\documentstyle[preprint,aps]{revtex}
\tighten
\begin{document}
\draft
\preprint{\vbox{Submitted to Phys.\ Rev.\ C \hfill FSU-SCRI-94-01 \\
                                       \null\hfill IU/NTC 93-29  \\
                                       \null\hfill nucl-th/9401001}}
\title{Relativistically generated asymmetry in the missing \\
       momentum distribution from the $(e,e'p)$ reaction}
\author{S. Gardner}
\address{Nuclear Theory Center and Department of Physics, \\
         Indiana University, Bloomington, IN 47505}
\author{J. Piekarewicz}
\address{Supercomputer Computations Research Institute, \\
         Florida State University, Tallahassee, FL 32306}
\date{\today}
\maketitle

\begin{abstract}
We calculate the asymmetry in the missing-momentum distribution
from the $(e,e'p)$ reaction in a relativistic formalism.
Longitudinal and transverse response functions are evaluated in
parallel kinematics as a function of the three-momentum transfer
to the nucleus. Analytic expressions for the responses
are obtained in a relativistic plane-wave impulse approximation.
These expressions reveal a large asymmetry in the momentum
distribution, at the plane-wave level, induced by the lower components
of the bound-state wave functions. These relativistic effects
contaminate any attempt to infer color transparency
from a measurement of the asymmetry in the $(e,e'p)$ reaction.
\end{abstract}

\narrowtext

\section{Introduction}
\label{secintro}

The $(e,e'p)$ reaction constitutes an invaluable tool in the study
of many diverse nuclear phenomena. The experimental setup enables
one to determine the 4-momentum of both the virtual photon and the
outgoing proton, so that the reaction kinematics are completely
determined. As a result, it was originally believed that the $(e,e'p)$
cross section provides direct information about the nucleon momentum
distribution in the nuclear medium~\cite{frumou84}. A plane-wave,
independent-particle description of the process suggests that the
$(e,e'p)$ cross section should be given by a set of discrete peaks
-- each corresponding to knock-out from a given nuclear shell -- with
the momentum distribution obtained by integrating the strength under
these peaks. This description is nearly adequate for states near the
Fermi surface, yet additional study has revealed that the
single-particle strength is strongly fragmented for deeply bound
states~\cite{quint86,herder88}. Knock-out from states near the Fermi
surface does support the notion of unfragmented single-particle strength,
but the occupancy of these states is less than one~\cite{quint86}.
These results show conclusively that the traditional picture of the
nucleus, as a collection of particles moving independently in a
mean-field potential, is too naive. Consequently, the $(e,e'p)$
reaction shows the limitations of an independent-particle
picture, yet it provides a fruitful testing ground for various
dynamical mechanisms, such as short-range correlations, that go beyond
a simple mean-field approach.

The $(e,e'p)$ reaction has also been used to study possible
medium modifications to the electromagnetic coupling of the
nucleon~\cite{steen86}. That such effects exist has been
suggested by a variety of studies. The original
European-Muon-Collaboration (EMC) effect, for example, can be
explained by a softening of the quark momentum distribution
in the nuclear medium~\cite{close85}, which, in turn, modifies
the nucleon's electromagnetic coupling.

Interest in the $(e,e'p)$ reaction has been recently stimulated by
suggestions that color transparency may manifest itself as an
asymmetry~\cite{jenkop93,nikol93,bianc93} in the missing-momentum
distribution, even in the absence of a modification of the total
cross section \cite{jenkop93}. ``Color transparency'' describes the
hypothesis that hadrons produced with large laboratory momenta
in certain exclusive reactions with nuclear targets interact in
an anomalously weak manner with the residual
nucleus~\cite{brodsky82,mueller82}.
We shall not concern ourselves directly with color transparency, yet
part of our interest in considering conventional relativistic effects
in the $(e,e'p)$ reaction is to the end of establishing a robust
``baseline'' calculation, against which possible novel effects may be
inferred.

The possible existence of color transparency has been discussed in a
great many different exclusive processes, yet the focus on the NE18
experiment at SLAC has made the quasielastic $(e,e'p)$ reaction the
paradigm~\cite{frank92}. With the advent of CEBAF, the focus on this
particular reaction is unlikely to diminish. Indeed, several
$(e,e'p)$ experiments have been approved in the hope of
identifying this novel behavior~\cite{saha91,milner91,gees91}.

Our purpose is to explore the asymmetry in the missing momentum
distribution in a conventional $(e,e'p)$ calculation. For
concreteness, let us now quantify what we mean by the ``asymmetry.''
We consider the $(e,e'p)$ reaction in the kinematics where
the three-momentum transfer ${\bf q}$ and the transverse missing
momentum $p_t$ are fixed (in all that follows, ``longitudinal''
and ``transverse'' are defined as the components parallel and
perpendicular to ${\bf q}$, respectively). With this choice of
kinematics, the longitudinal momentum distribution of the bound
proton is scanned by varying the energy transfer to the nucleus --
this results in a varying momentum for the outgoing proton. The
asymmetry is, then, defined as the integrated sum of events with
positive longitudinal missing momentum relative to those events
having negative missing momentum~\cite{bianc93}. Here we shall
consider the reaction specifically in ``parallel kinematics'', that
is, we shall take $p_t=0$, so that the knocked-out proton's momentum
is parallel to ${\bf q}$, in all that follows.

Studies of the $(e,e'p)$ reaction at low momentum transfers ($|{\bf q}|
{\hbox{${\lower.40ex\hbox{$<$}\atop\raise.20ex\hbox{$\sim$}}$}}1\;
{\rm GeV}$) yield missing-momentum ($p_{m}$) distributions
which are asymmetric about the $p_m=0$ point~\cite{bern82}. A
calculation in a nonrelativistic plane-wave impulse approximation
has no such asymmetry. The asymmetry at low momentum transfers is
understood to arise from the momentum dependence of the distortions
that the struck proton suffers in its exit from the
nucleus~\cite{ryck89}. In the momentum transfer regime beyond
$|{\bf q}| \sim 1\;{\rm GeV}$, nuclear quasielastic electron scattering
is essentially unexplored; this frontier provides a non-trivial testing
ground for phenomenologies constructed in a lower energy regime. At large
momenta, Glauber theory has been used to define the ``baseline''.
The attenuation of the ejected proton through the nucleus is given in
this model by the inelastic proton-nucleon cross section. For proton
momenta in excess of 3 GeV, the inelastic cross section is approximately
constant, so that the asymmetry is essentially zero.

The virtue of the asymmetry as a signal of transparency rests
wholly on the assumption that such an asymmetry is absent in
any conventional (i.e., without transparency) calculation of
the $(e,e'p)$ reaction at large momentum transfers. We shall show
that this assumption is flawed. Relativistic effects, even in a
plane-wave impulse approximation, serve to generate an asymmetry
in the $(e,e'p)$ reaction. In a previous work, one of us examined
the role of relativity on the spectral function~\cite{piereg92}.
In that context, relativistic effects, driven by particle-antiparticle
mixing, are responsible for generating features, such as a fragmentation
of the single-particle strength, which are associated nonrelativistically
with physics beyond the mean-field level. These relativistic effects exist
even if a mean-field ground state is used. As relativity generates
effects similar to what a more complicated nonrelativistic calculation
would generate, they can not be regarded as a definitive relativistic
signature. Here we consider an effect which is absent in the conventional
nonrelativistic formalism at the plane-wave level. Moreover, including
distortions as in the Glauber theory also results in a small asymmetry
at large momentum transfers. In contrast, relativistic corrections increase
with the momentum transfer to the nucleus and can become large. As these
effects are absent in the conventional nonrelativistic theory, they are
interesting in their own right. Moreover, their existence limits the
utility of the missing-momentum asymmetry as a signal of
color transparency~\cite{bianc93}.

We have organized the paper as follows. In Sec.~{\ref{secformal}}
we review some general properties of the $(e,e'p)$ reaction and
develop analytic expressions for the nuclear response in a relativistic
plane-wave impulse approximation. These plane-wave expressions identify
the lower components of the bound-state spinors as the source of the
asymmetry in the missing-momentum distribution. The only dynamical
input required is the relativistic bound state wave function.
Consequently, in Sec.~{\ref{secres}}, we use a ${}^{16}$O
ground-state wave function obtained in a relativistic mean-field
approximation to the Walecka model as an illustrative example.
Finally, we offer our conclusions in Sec~{\ref{secconcl}}.

\section{Formalism}
\label{secformal}

We begin by writing the inclusive hadronic tensor $W^{\mu\nu}(q)$ as
\begin{equation}
   W^{\mu\nu}(q) = \sum_{n}
   \langle \Psi_{n} | J^{\mu}(q) | \Psi_{0} \rangle
   \langle \Psi_{n} | J^{\nu}(q) | \Psi_{0} \rangle^{*}
   \delta(\omega-\omega_{n}) \;.
 \label{wincl}
\end{equation}
The nuclear electromagnetic current $J^{\mu}$ is responsible for
inducing transitions between the exact nuclear ground state
($\Psi_{0}$) and an excited state ($\Psi_{n}$) with excitation
energy $\omega_{n}=(E_{n}-E_{0})$. In an independent-particle
description of the ground state, the inclusive hadronic tensor
can be expressed in terms of individual particle-hole transitions;
i.e.,
\begin{equation}
    W^{\mu\nu}(q) = \sum_{\alpha\beta}
      W^{\mu\nu}({\bf q}; \beta\alpha)
      \delta(\omega-E_{\beta\alpha})       \;,
\end{equation}
where
\begin{equation}
   W^{\mu\nu}({\bf q}; \beta\alpha) \equiv
      J^{\mu}({\bf q}; \beta\alpha)
      J^{\nu\star}({\bf q}; \beta\alpha)   \;.
 \label{winclsp}
\end{equation}
Note that $\omega$ is the energy transfer to the nucleus, and
$E_{\beta\alpha}=(E_{\beta}-E_{\alpha})$ is the excitation energy
from the initial single-particle state $\alpha$ to the final state
$\beta$. The dynamical information about the process is contained in
\begin{equation}
     J^{\mu}({\bf q}; \beta\alpha) =
     \int d{\bf x} \;
     \overline{{\cal U}}_{\beta}({\bf x})
      e^{i{\bf q}\cdot{\bf x}} \; j^{\mu}(q) \;
       {\cal U}_{\alpha}({\bf x}) \;,
 \label{winclba}
\end{equation}
the transition matrix element between an occupied single-particle
state ${\cal U}_{\alpha}$ and an unoccupied state ${\cal U}_{\beta}$.
We assume the impulse approximation valid and employ an electromagnetic
current with single-nucleon form factors parameterized from on-shell
$eN$ scattering. That is,
\begin{equation}
   j^{\mu}(q) = F_{1}(Q^{2}) \gamma^{\mu} +
                F_{2}(Q^{2}) i\sigma^{\mu\nu} {q_{\nu} \over 2M} \;,
                \quad (Q^{2} \equiv {\bf q}^{2}-\omega^{2})
 \label{nucem}
\end{equation}
where $F_1(Q^2)$ and $F_2(Q^2)$ are given as in Ref.~\cite{horpie93}.
Specifically, for the proton,
\begin{mathletters}
 \label{emformf}
  \begin{eqnarray}
   F_{1}(Q^{2}) &=&
    \left[1+\tau(1+\kappa_{p})\right]
    (1+4.97\tau)^{-2} / (1+\tau)                     \;, \\
   F_{2}(Q^{2}) &=&
    \kappa_{p}
    (1+4.97\tau)^{-2} / (1+\tau)                     \;.
  \end{eqnarray}
\end{mathletters}
Note that $\tau \equiv Q^{2}/4M^2$, and $\kappa_{p}=1.793$ is
the proton's anomalous magnetic moment.

\subsection{Relativistic plane-wave impulse approximation}
\label{rpwia}

	The plane-wave limit of the $(e,e'p)$ reaction is
useful in providing crude guidance into the nature of the
reaction. Indeed, the plane-wave impulse approximation
(PWIA) analysis leads one to believe that the $(e,e'p)$
reaction can be used to extract the spectral function
and, thus, the nucleon momentum distribution~\cite{frumou84}.
This is only approximately correct. For example,
final-state interactions (FSI) between the outgoing nucleon and
the residual nucleus are known to modify the plane-wave picture.
Of particular relevance to our present work is the symmetry in the
missing-momentum distribution predicted in the nonrelativistic PWIA.
This symmetry is broken by final-state interactions. Yet, the inclusion
of distortions in the Glauber formalism suggests that the symmetry in
the missing-momentum distribution will become approximately restored at
large momentum transfers. This observation supports the claim made in
Ref.~\cite{bianc93} that any measured asymmetry in the missing-momentum
distribution is due to the formation of a compact state and, thus,
constitutes evidence in support of color transparency.

In this section we will show that some notions developed
in a nonrelativistic context no longer hold in a relativistic
formalism. In particular, we will prove that the symmetry in the
missing-momentum distribution is broken at the plane-wave level
due to relativistic effects. Moreover, the asymmetry increases
with the momentum transferred to the nucleus -- and as the momentum
of the outgoing nucleon -- is increased. Consequently, these
relativistic ``corrections'' to the asymmetry contaminate any
attempt to identify color transparency.

In a relativistic plane-wave impulse approximation (RPWIA) the
outgoing proton with momentum ${\bf p'}$ and spin projection $s'$
is described by a plane-wave Dirac spinor~\cite{bjodre65}:
\begin{eqnarray}
 {\cal U}_{{\bf p'}s'}({\bf x}) &=&
  e^{i{\bf p'}\cdot{\bf x}} \;
 {\cal U}({\bf p'},s') \;,        \\
 {\cal U}({\bf p'},s')           &=&
 \sqrt{{E_{{\bf p'}} + M \over 2E_{{\bf p'}}}}
 \left(
  \begin{array}{c}
            1      \\
      \displaystyle{
        {\sigma\cdot{\bf p'} \over
         E_{{\bf p'}} + M }}
  \end{array}
 \right) \chi_{s'} \;,
 \label{freewf}
\end{eqnarray}
where the on-shell energy of the outgoing proton is given by
\begin{equation}
 E_{\bf p'}=\sqrt{{\bf p'}^{2}+M^{2}} \;.
\end{equation}
Note that the normalization implied by Eq.~(\ref{freewf}) differs
from the one in Ref.~\cite{bjodre65} in that
\begin{equation}
  {\cal U}^{\dagger}({\bf p},s)     \;
  {\cal U}({\bf p},s')=\delta_{ss'} \;.
\end{equation}
In the following presentation, we shall use only the Dirac
(or vector) piece of the electromagnetic current, for simplicity.
Our final results, presented at the end of this section, include
the anomalous piece of the current as well.

In a RPWIA the Dirac piece of the current matrix element for the
knock-out of a bound-state nucleon is
\begin{equation}
     J^{\mu}_{\scriptscriptstyle D}({\bf q},{\bf p'},s';\alpha) =
       \overline{\cal U}({\bf p'},s')
       \gamma^{\mu}{\cal U}_{\alpha}({\bf p}) \;,
 \label{jmupwia}
\end{equation}
where ${\cal U}({\bf p'} s')$ is the plane-wave state defined
above and where
the Fourier transform of the relativistic
bound-state wave function is given by
\begin{equation}
  {\cal U}_{\alpha}({\bf p}) = \int d{\bf x} \;
      e^{-i{\bf p}\cdot{\bf x}}  \;
      {\cal U}_{\alpha}({\bf x}) \;.
 \label{uofp}
\end{equation}
Note that we have also introduced the missing momentum
\begin{equation}
   {\bf p}_{m} \equiv {\bf p} = {\bf p}^{\prime} - {\bf q} \;.
 \label{pmiss}
\end{equation}
The semi-inclusive hadronic tensor can now be written in terms of
a Feynman trace,
\begin{eqnarray}
  W^{\mu\nu}_{\scriptscriptstyle DD}({\bf q},{\bf p'},s';\alpha) &=&
    \left[
      \overline{\cal U}({\bf p'},s')
      \gamma^{\mu}{\cal U}_{\alpha}({\bf p})
    \right]
    \left[
      \overline{\cal U}({\bf p'},s')
      \gamma^{\nu}{\cal U}_{\alpha}({\bf p})
    \right]^{\star}                  \nonumber   \\   &=&
    {\rm Tr}
    \left[
      \gamma^{\mu} S_{\alpha}({\bf p}) \gamma^{\nu}
      \Big({{\rlap/{p'}} + M \over 2E_{{\bf p'}}}\Big)
      \Big({1+\gamma^{5}{\rlap/{s'}} \over 2}\Big)
    \right] \;,
 \label{trace}
\end{eqnarray}
involving the spin-dependent projection operator for the
outgoing nucleon~\cite{bjodre65}, as well as the bound-state
``propagator''
\begin{equation}
  S_{\alpha}({\bf p}) =
        {\cal U}_{\alpha}({\bf p}) \;
        \overline{\cal U}_{\alpha}({\bf p}) \;,
 \label{salpha}
\end{equation}
which satisfies the normalization condition
\begin{equation}
  \int {d{\bf p} \over (2\pi)^{3}}
    {\rm Tr}\Big[\gamma^{0}S_{\alpha}({\bf p})\Big]=1 \;.
 \label{normal}
\end{equation}

To proceed, we need an explicit form for $S_{\alpha}({\bf p})$.
To do so, we recognize that in the presence of a spherically
symmetric potential the eigenstates of the Dirac equation can be
classified according to a generalized angular momentum $\kappa$
and can be written in a two component representation; i.e.,
\begin{equation}
  {\cal U}_{E\kappa m}({\bf x})= {1 \over x}
   \left[
    \begin{array}{c}
      \phantom{i}
       g_{E\kappa}(x)
      {\cal{Y}}_{+\kappa m}(\hat{\bf{x}})    \\
       i
       f_{E\kappa}(x)
      {\cal{Y}}_{-\kappa m}(\hat{\bf{x}})
     \end{array}
   \right] \;.
 \label{dirwfx}
\end{equation}
The upper and lower components are expressed in terms of
spin-spherical harmonics defined by
\begin{equation}
  {\cal{Y}}_{\kappa m}(\hat{\bf{x}}) \equiv
  \langle\hat{\bf{x}}|l{\scriptstyle{1\over 2}}jm\rangle \;; \quad
  j=|\kappa|-{1\over 2} \;; \quad
  l=\cases{   \kappa\;,   & if $\kappa > 0$; \cr
           -1-\kappa\;,   & if $\kappa < 0$. \cr}
 \label{curlyy}
\end{equation}
The Fourier transform of the above bound-state wave function
can now be easily evaluated. That is,
\begin{equation}
  {\cal U}_{E\kappa m}({\bf p}) =
  {4\pi \over p} (-i)^{l}
   \left[
    \begin{array}{c}
       \phantom{-}
       g_{E\kappa}(p)
      {\cal{Y}}_{+\kappa m}(\hat{\bf{p}})  \\
      -f_{E\kappa}(p)
      {\cal{Y}}_{-\kappa m}(\hat{\bf{p}})
     \end{array}
   \right] =
 {4\pi \over p} (-i)^{l}
   \left[
    \begin{array}{c}
       g_{E\kappa}(p) \\
       f_{E\kappa}(p) ({\bf \sigma}\cdot{\hat{\bf p}})
     \end{array}
   \right] {\cal{Y}}_{+\kappa m}(\hat{\bf{p}}) \;,
 \label{dirwfp}
\end{equation}
where we have written the Fourier transforms of the radial
wave functions as
\begin{mathletters}
\begin{eqnarray}
   g_{E\kappa}(p) &=& \int_{0}^{\infty} dx \;
    g_{E\kappa}(x)
    \hat{\jmath}{\hbox{\lower 3pt\hbox{$_l$}}}(px) \;, \\
   f_{E\kappa}(p) &=& ({\rm sgn}\kappa) \int_{0}^{\infty} dx  \;
    f_{E\kappa}(x)
    \hat{\jmath}{\hbox{\lower 3pt\hbox{$_{l'}$}}}(px) \;.
\end{eqnarray}
\label{gfp}
\end{mathletters}
In the above expression we have introduced
$\hat{\jmath}{\hbox{\lower 3pt\hbox{$_{l}$}}}(z)$,
the Riccati-Bessel function, through the relation
$\hat{\jmath}{\hbox{\lower 3pt\hbox{$_{l}$}}}(z)
=zj{\hbox{\lower 3pt\hbox{$_{l}$}}}(z)$.
By now making use of the following identities
\begin{mathletters}
\begin{eqnarray}
  \sum_{m}
   {\cal{Y}}_{+\kappa m}(\hat{\bf{p}})
   {\cal{Y}}^{\star}_{+\kappa m}(\hat{\bf{p}})  &=&
  \sum_{m}
   {\cal{Y}}_{-\kappa m}(\hat{\bf{p}})
   {\cal{Y}}^{\star}_{-\kappa m}(\hat{\bf{p}})  =
  {2j+1 \over 8\pi} \;,                     \\
  \sum_{m}
   {\cal{Y}}_{+\kappa m}(\hat{\bf{p}})
   {\cal{Y}}^{\star}_{-\kappa m}(\hat{\bf{p}})  &=&
  \sum_{m}
   {\cal{Y}}_{-\kappa m}(\hat{\bf{p}})
   {\cal{Y}}^{\star}_{+\kappa m}(\hat{\bf{p}})  =
  -{2j+1 \over 8\pi}\;({\bf \sigma}\cdot\hat{\bf p}) \;,
\end{eqnarray}
\label{msum}
\end{mathletters}
we can express the bound-state propagator in closed form.
That is,
\begin{equation}
  S_{\alpha}({\bf p}) \equiv S_{E\kappa}({\bf p}) =
  {1 \over 2j+1} \sum_{m}
     {\cal U}_{E\kappa m}({\bf p}) \;
     \overline{\cal U}_{E\kappa m}({\bf p}) =
     ({\rlap/{p}}_{\alpha} + M_{\alpha})    \;,
 \label{sek}
\end{equation}
where we have defined the above mass-, energy-, and momentum-like
quantities as
\begin{mathletters}
\begin{eqnarray}
  M_{\alpha} &=& \left({\pi \over p^{2}}\right)
                  \Big[g_{\alpha}^{2}(p) -
                       f_{\alpha}^{2}(p)\Big] \;, \\
  E_{\alpha} &=& \left({\pi \over p^{2}}\right)
                  \Big[g_{\alpha}^{2}(p) +
                       f_{\alpha}^{2}(p)\Big] \;, \\
  {\bf p}_{\alpha} &=& \left({\pi \over p^{2}}\right)
                   \Big[2 g_{\alpha}(p)
                          f_{\alpha}(p)\hat{\bf p}
                   \Big]  \;.
 \label{epm}
\end{eqnarray}
\end{mathletters}
Note that they satisfy the ``on-shell relation''
\begin{equation}
  p_{\alpha}^{2}=E_{\alpha}^{2}-{\bf p}_{\alpha}^{2}
                =M_{\alpha}^{2} \;.
 \label{onshell}
\end{equation}
The above representation of the bound-state propagator,
given entirely in terms of Dirac gamma matrices, enables
one to evaluate the semi-inclusive hadronic tensor,
\begin{eqnarray}
    W^{\mu\nu}_{\scriptscriptstyle DD}({\bf q};{\bf p'}s'\alpha) =
    {\rm Tr}
    \left[
      \gamma^{\mu} ({\rlap/{p}}_{\alpha} + M_{\alpha}) \gamma^{\nu}
      \Big({{\rlap/{p'}} + M \over 2E_{{\bf p'}}}\Big)
      \Big({1+\gamma^{5}{\rlap/{s'}} \over 2}\Big)
    \right] \;,
 \label{feyntrace}
\end{eqnarray}
with Feynman's trace techniques.
This relativistic plane-wave expression for the Dirac component of
the hadronic tensor is completely general and does not assume any
specific kinematical constraint. In the present work, however, we
are interested in the missing-momentum asymmetry evaluated in
parallel kinematics ($\hat{\bf p}'\cdot\hat{\bf q}\!=\!1$).
For this case, then, the charge, transverse, and longitudinal
components of the hadronic tensor are given, respectively, by
\begin{eqnarray}
  W^{00}_{\scriptscriptstyle DD}({\bf q},{\bf p'};\alpha) &=&
    \left({E_{\bf p'} + M \over 2E_{\bf p'}}\right)
    \left({4\pi \over p^{2}}\right)
    \Big[g_{\alpha}(p) +
    (\hat{\bf p}\cdot\hat{\bf p}') \;
     \xi_{\bf p'}f_{\alpha}(p)\Big]^{2} \;, \\
  W^{11}_{\scriptscriptstyle DD}({\bf q},{\bf p'};\alpha) &=&
    \left({E_{\bf p'} + M \over 2E_{\bf p'}}\right)
    \left({4\pi \over p^{2}}\right)
    \Big[\xi_{\bf p'}g_{\alpha}(p) -
    (\hat{\bf p}\cdot\hat{\bf p}')
      f_{\alpha}(p)\Big]^{2} \;, \\
  W^{33}_{\scriptscriptstyle DD}({\bf q},{\bf p'};\alpha) &=&
    \left({E_{\bf p'} + M \over 2E_{\bf p'}}\right)
    \left({4\pi \over p^{2}}\right)
    \Big[\xi_{\bf p'}g_{\alpha}(p) +
    (\hat{\bf p}\cdot\hat{\bf p}')
    f_{\alpha}(p)\Big]^{2} \;,
 \label{ws}
\end{eqnarray}
where we have defined
\begin{equation}
\xi_{\bf p'} \equiv {|{\bf p'}| \over{ (E_{{\bf p'}} + M)}} \;.
\end{equation}
Two aspects of these results are particularly noteworthy.
First, we have included the charge as well as the longitudinal
components of the hadronic tensor because the electromagnetic
current is not conserved at the plane-wave level. Second, the
term $(\hat{\bf p}\cdot\hat{\bf p}')$ generates an asymmetry in the
missing-momentum distribution. This has its origin in the relation
\begin{equation}
  \hat{\bf p}\cdot\hat{\bf p}'=
  \hat{\bf p}\cdot\hat{\bf q }=
  \cases{
          +1\;, & if $|{\bf p'}| > |{\bf q}| \;;$  \cr
          -1\;, & if $|{\bf p'}| < |{\bf q}| \;.$  }
\end{equation}
Note, however, that in the absence of lower bound-state components
(i.e., $f_{\alpha}\equiv 0$), there is no asymmetry, in agreement
with nonrelativistic expectations. Moreover, including distortions
as in the Glauber theory does not significantly alter this result
at large momentum transfers. The asymmetry at large momenta, then,
is a genuinely relativistic effect. This simple result, namely, the
relativistic generation of an asymmetry at the plane-wave level, is
the main result of the present work.

One of the advantages of plane-wave analyses is that they
suggest that the $(e,e'p)$ cross section can be factorized
into an (off-shell) elementary electron-proton cross section
times a spectral function containing all nuclear-structure
information. In order to isolate the nuclear structure effects,
we shall introduce a reduced cross section. This reduced cross
section is obtained from the full $(e,e'p)$ cross section by
dividing out an off-shell electron-proton cross section -- with
some kinematical factors. Unfortunately, the off-shell
electron-proton cross section is not uniquely defined~\cite{defor83}.
The choice made can even obscure the dynamical features that one is
trying to uncover. Indeed, an off-shell electron-proton cross section
also generates an asymmetry in the missing-momentum distribution, but
this asymmetry is unrelated to nuclear-structure effects. For our present
purpose, we shall adopt the following procedure. First, we separate the
full cross section into longitudinal and transverse response functions.
Then, we divide each of these responses by an off-shell electric or
magnetic single-nucleon form factor, as appropriate, to yield the reduced
cross sections. This definition has the advantage that the nonrelativistic
limit of the reduced cross section does, indeed, yield a symmetric
missing-momentum distribution. This procedure provides, then, a baseline
for comparing different theoretical mechanisms of breaking the symmetry,
be they relativistic effects, or color transparency.

We shall now introduce longitudinal and transverse RPWIA reduced cross
sections for the $(e,e'p)$ reaction. The reduced cross sections
have been computed with the full electromagnetic current -- including
the anomalous piece -- and are given by
\begin{mathletters}
  \label{rholt}
\begin{eqnarray}
  \hbox{\raise.8ex\hbox{$\rho$}\kern-2pt}_{L}({\bf q},{\bf p'};\alpha)
   &\equiv&
   {W^{00}({\bf q},{\bf p'};\alpha) \over
    w^{00}({\bf q},{\bf p'})} =
    \left({4\pi \over p^{2}}\right) g_{\alpha}^{2}(p)
    \left[
        1 \pm
        \left({\xi_{\bf p'}F_{1}+\bar{q}F_{2} \over
               F_{1}-\xi_{\bf p'}\bar{q}F_{2}}\right)
               {f_{\alpha}(p) \over g_{\alpha}(p)}
    \right]^{2} \;,  \label{rhol} \\
  \hbox{\raise.8ex\hbox{$\rho$}\kern-2pt}_{T}({\bf q},{\bf p'};\alpha)
   &\equiv&
   {W^{11}({\bf q},{\bf p'};\alpha) \over
    w^{11}({\bf q},{\bf p'})} =
    \left({4\pi \over p^{2}}\right) g_{\alpha}^{2}(p)
    \left[
        1 \mp
        \left({F_{1}+(\bar{\omega}-\xi_{\bf p'}\bar{q})F_{2} \over
        \xi_{\bf p'}F_{1}+
        (\bar{q}-\xi_{\bf p'}\bar{\omega})F_{2}}\right)
        {f_{\alpha}(p) \over g_{\alpha}(p)}
    \right]^{2} \;.
 \label{rhot}
\end{eqnarray}
\end{mathletters}
The off-shell single-nucleon responses, $w^{00}$ and $w^{11}$, have
been chosen as
\begin{mathletters}
  \label{slstfree}
\begin{eqnarray}
  w^{00}({\bf q},{\bf p'}) &=&
    \left({E_{\bf p'} + M \over 2E_{\bf p'}}\right)
    \Big[
       F_{1}-\xi_{\bf p'}\bar{q}F_{2}
    \Big]^{2} \longrightarrow
    \left({1 +\tau \over 1+2\tau}\right)G_{E}^{2}(Q^{2}) \;,
    \label{slfree} \\
  w^{11}({\bf q},{\bf p'}) &=&
    \left({E_{\bf p'} + M \over 2E_{\bf p'}}\right)
    \Big[
       \xi_{\bf p'}F_{1}+
       (\bar{q}-\xi_{\bf p'}\bar{\omega})F_{2}
    \Big]^{2} \longrightarrow
    \left({\tau \over 1+2\tau}\right)G_{M}^{2}(Q^{2}) \;,
    \label{stfree}
\end{eqnarray}
\end{mathletters}
where the following dimensionless quantities have been introduced
\begin{equation}
 \tau \equiv {Q^2 \over 4M^{2}}        \;, \quad
 \bar{q} \equiv {|{\bf q}| \over 2M}   \;, \quad
 \bar{\omega} \equiv {\omega \over 2M} \;.
\end{equation}
In constructing the single-nucleon responses we have demanded
that the resulting missing-momentum distribution be symmetric
at the plane-wave level in the limit of a vanishing lower
bound-state component. Note that in this limit
\begin{mathletters}
 \begin{eqnarray}
   W^{00}({\bf q},{\bf p'};\alpha)|_{f_{\alpha}=0} &=&
  {4\pi \over {p^2}} g_{\alpha}^2(p)w^{00}({\bf q},{\bf p'}) \;, \\
   W^{11}({\bf q},{\bf p'};\alpha)|_{f_{\alpha}=0} &=&
  {4\pi \over {p^2}} g_{\alpha}^2(p)w^{11}({\bf q},{\bf p'}) \;.
 \end{eqnarray}
\end{mathletters}
The arrow in the previous expressions is used to indicate
that the single-nucleon responses in the isolated nucleon
limit (so that $p=0$) become proportional to the electric
and magnetic form factors of the nucleon. The off-shell
single-nucleon responses have been defined to the
end of establishing a baseline calculation for the asymmetry,
against which relativistic effects may be assessed.

The longitudinal reduced cross section clearly shows
that the symmetry in the missing momentum distribution
is broken by relativistic effects. Consequently, the
symmetry is restored when all momenta become small
relative to the nucleon mass. The nonrelativistic limit
of the transverse reduced cross section is, in contrast,
not clearly defined. The relativistic ``corrections'' are
of ${\cal O}(1)$, so that the transverse response is an
intrinsically relativistic observable.

In a recent publication, Bianconi, Boffi, and Kharzeev
suggested~\cite{bianc93} studying the asymmetry in the
missing-momentum distribution as a means of uncovering
color transparency. In particular, they suggest comparing
the number of events with positive missing momentum
($|{\bf p'}|>|{\bf q}|$) to those events with negative
missing momentum ($|{\bf p'}|<|{\bf q}|$)~\cite{bianc93}.
In this work we follow their suggestion and define the
following asymmetries. Here we define, however, separate
longitudinal and transverse asymmetries:
\begin{mathletters}
 \label{asymm}
 \begin{eqnarray}
  A_{L}({\bf q}, p_{\rm max}) &\equiv&
   {N_{L}^{(+)}({\bf q}, p_{\rm max})
- N_{L}^{(-)}({\bf q}, p_{\rm max}) \over
    N_{L}^{(+)}({\bf q}, p_{\rm max})
+ N_{L}^{(-)}({\bf q}, p_{\rm max})}
   \mathop{\longrightarrow}\limits_{f_{\alpha}\equiv 0} 0 \;,
   \label{asymml} \\
  A_{T}({\bf q}, p_{\rm max}) &\equiv&
   {N_{T}^{(+)}({\bf q}, p_{\rm max})
- N_{T}^{(-)}({\bf q}, p_{\rm max}) \over
    N_{T}^{(+)}({\bf q}, p_{\rm max})
+ N_{T}^{(-)}({\bf q}, p_{\rm max})}
   \mathop{\longrightarrow}\limits_{f_{\alpha}\equiv 0} 0 \;.
   \label{asymmt}
 \end{eqnarray}
\end{mathletters}
In all that follows, we will assume that we are studying
single-particle transitions from a specific bound state
$\alpha$ to the continuum, so that the $\alpha$ label is
suppressed -- both in the above and in the reduced cross
sections. The asymmetries are obtained from the total number
of longitudinal or transverse events having either positive
or negative missing momentum:
\begin{mathletters}
 \label{pmint}
 \begin{eqnarray}
  N_{L,T}^{(+)}({\bf q}, p_{\rm max}) &\equiv&
   \int_{0}^{p_{\rm max}} {p^{2}d{p} \over 2\pi^{2}}
   \hbox{\raise.8ex\hbox{$\rho$}\kern-2pt}_{L,T}({\bf q},{\bf p'})
   \mathop{\longrightarrow}\limits_{f_{\alpha}\equiv 0}
   \left({2 \over \pi}\right)
   \int_{0}^{p_{\rm max}} d{p} \; g_{\alpha}^{2}(p)
   \mathop{\longrightarrow}\limits_{p_{\rm max}\rightarrow\infty}
 1 \;, \\
  N_{L,T}^{(-)}({\bf q}, p_{\rm max}) &\equiv&
   \int_{-p_{\rm max}}^{0} {p^{2}d{p} \over 2\pi^{2}}
   \hbox{\raise.8ex\hbox{$\rho$}\kern-2pt}_{L,T}({\bf q},{\bf p'})
   \mathop{\longrightarrow}\limits_{f_{\alpha}\equiv 0}
   \left({2 \over \pi}\right)
   \int_{-p_{\rm max}}^{0} d{p} \; g_{\alpha}^{2}(p)
   \mathop{\longrightarrow}\limits_{p_{\rm max}\rightarrow\infty}
 1 \;.
 \end{eqnarray}
\end{mathletters}
The arrows indicate two successive limits. In the first
the lower bound-state component is set to zero, while
in the second the cut-off $p_{\rm max}$ goes to infinity.
Introducing a cut-off is useful as it enables us to exclude
those large missing momentum events that might be sensitive
to physics beyond the mean-field picture. Note that when
$f_{\alpha}=0$ the number of events with positive and
negative missing momenta become equal, so that the asymmetry
vanishes. The symmetry is broken, however, by the inclusion
of momentum-dependent final-state interactions -- or by
relativistic effects.

The symmetry may also be broken by color transparency~\cite{jenkop93}.
Color transparency, if it exists, is due to the coherent interference
of the distortions suffered by the outgoing proton and other, higher-mass,
baryon states produced by the virtual photon -- the strong interactions
of the individual components conspire to cancel, at least in part. The
initial and final electron 4-momenta fix $x_{\rm B}$
($x_{\rm B}\equiv~Q^2/2M\omega$) so that
the longitudinal component of the struck proton momentum depends on
whether a proton or a $N^*$ is produced. For $x_{\rm B}> 1$, the
bound-state momenta $p$ required to make a proton ($N$) or an excited
baryon state ($N^*$) must be of the same sign. Satisfying the kinematic
constraints for both $N$ and $N^*$ production at moderate momentum
transfers is unlikely, as bound-nucleon momenta in the tail of the
momentum distribution are required. For $x_{\rm B}< 1$, the required
momenta can be of opposite sign, so that the kinematic constraints can
be satisfied at moderate $Q^2$ with a finite probability. This is the
source of the suggested asymmetry effect~\cite{jenkop93}. Bianconi, Boffi,
and Kharzeev~\cite{bianc93} relate the proposed asymmetry in $x_{\rm B}$
to an asymmetry in the missing-momentum distribution; they find a
positive asymmetry, as we shall find here for the longitudinal asymmetry.
However, their arguments give no insight into how the longitudinal and
transverse asymmetries may differ, nor is it helpful in disentangling
off-shell form factor effects from the modification of the final-state
interactions. Yet, the conventional Glauber analysis suggests that the
asymmetry should be small at large $Q^2$, supporting the asymmetry as a
signal of color transparency, as argued by the authors of
Ref.~{\cite{bianc93}}.

We believe that relativistic effects modify this picture. Indeed,
we have shown that relativistic effects generate an asymmetry in
the missing-momentum distribution at the plane-wave level.

\section{Results}
\label{secres}

We now proceed to show relativistic plane-wave results for the
missing-momentum asymmetry. In addition, we estimate how the
asymmetry is changed upon the inclusion of relativistic distortions
which are essential for restoring electromagnetic gauge invariance.
At the plane-wave level, the reduced cross sections, in Eq.~(\ref{rholt}),
require only the bound-state wave functions as dynamical input. We
have assumed  that the single-nucleon form factors are not modified by
the nuclear medium and have simply adopted the parameterization for
the on-shell form factors given in Eq.~(\ref{emformf}). The bound-state
wave functions have been obtained in a relativistic mean-field
approximation to the Walecka model~\cite{horser81}. In the Walecka
model nucleons interact via the exchange of neutral scalar ($\sigma$)
and vector ($\omega$) mesons~\cite{wal74,serwal86}. The parameters of
the model are adjusted, at the mean-field level, to reproduce the
binding energy and density of nuclear matter at saturation. With this
input, the mean-field approximation to the Walecka model has been
successful in describing the ground-state properties of finite
nuclei -- as well as the linear response of the ground state to
a variety of probes~\cite{serwal86}. This is realized, in part,
because the Walecka model is characterized by strong scalar and
(timelike) vector potentials that cancel to generate the observed
weak binding energies, but combine to yield a strong spin-orbit
splitting for the positive-energy states~\cite{serwal86}.

In Fig.~\ref{figzero} we present the charge response as a function
of the missing momentum for the two valence orbitals in ${}^{16}$O
for the three-momentum transfers of $|{\bf q}|=1,5~$GeV. Noting
Eq.~(\ref{pmint}a), the response has been multiplied by the
phase-space factors necessary to make the area under the curve
directly proportional to the number of integrated events. In addition,
we have scaled the response in order to compensate for the rapid
falloff of the form factors, since we have chosen to use the same
scale here as in Fig.~\ref{figone}. These constant factors do not
change the asymmetries -- these are displayed in brackets next to
the appropriate momentum-transfer labels.

The asymmetry in the charge response is driven by a combination
of nuclear-structure and single-nucleon effects. As we have
discussed, nuclear-structure effects generate an asymmetry in
the momentum distribution due to the presence of lower components
in the bound-state wave function. However, even in the absence of
relativistic corrections, the charge response, at fixed three-momentum
transfer, will display an asymmetry from single-nucleon effects.
The latter asymmetry is driven by the $Q^{2}$ dependence of
the form factor. The change in $Q^2$ itself depends on the magnitude
and on the direction of the missing momentum. For fixed $|{\bf p}|$,
the energy transfer to a nucleon moving parallel to $\hat{\bf q}$ must
be larger than that to a nucleon moving antiparallel to ${\bf q}$, in order
to yield an outgoing nucleon which is on its mass shell. Hence, the
electromagnetic coupling is stronger for the nucleon moving parallel
to ${\bf q}$ since its charge form factor is being probed at a lower
value of $Q^{2}$. This single-nucleon effect is responsible for generating
a positive asymmetry in the missing momentum distribution. Its contribution
must, then, be eliminated from the response before one can identify the role
of the nuclear medium in breaking the symmetry.

Eliminating the single-nucleon contribution from the asymmetry
is, however, not free from ambiguity. Indeed, certain off-shell
prescriptions could accentuate the effect. We believe that the
off-shell prescription adopted here [Eq.~(\ref{slfree})] is an
appealing choice for the study of the asymmetry. This choice yields,
in particular, a symmetric missing-momentum distribution
in the nonrelativistic plane-wave limit. This is an useful
baseline against which interesting behavior may be established.

In Fig.~\ref{figone} we present the longitudinal reduced cross
section [Eq.~(\ref{rhol})] for the two valence orbitals in
${}^{16}$O for the three-momentum transfers of $|{\bf q}|=1,5~$GeV.
In this case the area under the curve is directly proportional to
the number of events $N_{L}^{(\pm)}$ defined in Eq.~(\ref{pmint}).
The reduced cross section clearly shows an asymmetric distribution.
This asymmetry is entirely due to relativistic nuclear-structure
effects. The total asymmetries are displayed in the appropriate
brackets. In particular, we note that removing the single-nucleon
contribution to the response yields asymmetries which are slightly
smaller than those of Fig.~\ref{figzero}.

In Fig.~\ref{figtwo} we show the longitudinal asymmetry
[Eq.~(\ref{asymml})] as a function of the three-momentum
transfer $|{\bf q}|$ and the momentum cut-off $p_{\rm max}$.
At small three-momentum transfers, the longitudinal asymmetry
is negligible because the relativistic corrections are small
($\xi_{\bf p'}\simeq |{\bf p'}|/2M\ll 1$, and
$\bar{q}=|{\bf q}|/2M \ll 1$). In contrast, at large momentum
transfers, the relativistic ``corrections'' are large and so is the
asymmetry. We can understand this by examining the high-energy limit
of the longitudinal reduced cross section [Eq.~\ref{rhol}]. At large
three-momentum transfers ($\bar{q}=|{\bf q}|/2M \gg 1$) and small missing
momenta ($|{\bf p}|/2M \ll 1$), we obtain the following limits for
the other previously defined kinematical variables
\begin{mathletters}
\begin{eqnarray}
  \xi_{\bf p'} &\equiv& {|{\bf p'}| \over E_{{\bf p'}} + M}
  \rightarrow 1 \;, \\
  \omega &\equiv& E_{\bf p'} - E \rightarrow |{\bf p'}| - M \;, \\
  \tau   &\equiv& {Q^{2} \over 4M^{2}} \rightarrow \bar{q} \;.
\end{eqnarray}
\end{mathletters}
These relations imply, in particular, that the following ratio of
anomalous to Dirac form factors becomes
\begin{equation}
 \bar{q} \; {F_{2} \over F_{1}} =
 \bar{q} \; \left[{\kappa_{p} \over 1+\tau(1+\kappa_{p})}\right]
 \rightarrow
 {\kappa_{p} \over (1+\kappa_{p})} \;,
\end{equation}
so that one finds the following simple expression for the
reduced cross section:
\begin{equation}
  \hbox{\raise.8ex\hbox{$\rho$}\kern-2pt}_{L}({\bf q},{\bf p'};\alpha)
   \rightarrow
    \left({4\pi \over p^{2}}\right) g_{\alpha}^{2}(p)
    \left[
        1 \pm
        (1+2\kappa_{p})
        {f_{\alpha}(p) \over g_{\alpha}(p)}
    \right]^{2} \;.
\end{equation}
This result indicates that the longitudinal asymmetry is enhanced
not only by the in-medium wave function ratio of
($f_{\alpha}/g_{\alpha}$) but also by an additional factor
of [$(1+2\kappa_{p})\sim 4.6$].

In Fig.~\ref{figthree} we show, for completeness, the transverse
asymmetry. The transverse response is an intrinsically relativistic
observable, so that the transverse asymmetry is large even at small
momentum transfers. The asymmetry, however, saturates rapidly, as
reflected by the high-energy limit of the reduced cross section:
\begin{equation}
  \hbox{\raise.8ex\hbox{$\rho$}\kern-2pt}_{T}({\bf q},{\bf p'};\alpha)
   \rightarrow
    \left({4\pi \over p^{2}}\right) g_{\alpha}^{2}(p)
    \left[
        1 \mp
        {f_{\alpha}(p) \over g_{\alpha}(p)}
    \right]^{2} \;.
\end{equation}

We conclude this section with a brief discussion of distortion
effects. A nonrelativistic analysis of the $(e,e'p)$ reaction
generates an asymmetry in the missing-momentum distribution
when distortion effects are included, at low energies.
At high energies, however, a Glauber analysis suggests that the
symmetry should be restored. Relativistically, however, the
energy dependence of the distorting potential depends strongly
on its Lorentz character and can be substantially different from
its nonrelativistic counterpart~\cite{piekar93}.

Distortions play an important role in the conservation of the
electromagnetic current. Electromagnetic gauge invariance has
been lost as a result of our plane-wave treatment, and distortions
play a fundamental role in its restoration. The included distortions,
however, must be consistent with the mean-field potential used to
obtain the single-particle spectrum. Hence, in calculating the distorted
wave for the outgoing proton, we have employed the same real and
energy-independent potential used to generate the mean-field ground
state. Unfortunately, this treatment implies that the imaginary part
of the nucleon self-energy -- which is physically important -- has been
suppressed. In spite of this, we have calculated the nuclear response
in a relativistic distorted-wave impulse approximation (RDWIA), in order
to show what impact the restoration of gauge invariance has on the
asymmetry. As we shall see, the longitudinal asymmetry remains large in
the presence of the included distortions. The incorporation of absorptive
and energy-dependent effects in a calculation with electromagnetic gauge
invariance remains an important open problem.

In Fig.~\ref{figfour} we show the longitudinal reduced cross section
as a function of the missing momentum for the two valence orbitals
in ${}^{16}$O at $|{\bf q}|=1$~GeV. The dashed line is the relativistic
plane-wave result also shown in Fig.~\ref{figone}. The symmetric
distribution (dot-dashed line) results from neglecting the lower
component of the bound-state wave function in the RPWIA calculation.
This represents our best attempt in generating a nonrelativistic result.
Finally, relativistic distorted-wave results are shown by the solid
line. The positive shift in the RDWIA distribution relative to the
plane-wave result is due to the repulsive character of the real distorting
potential. The self-consistent, mean-field potential employed in the RDWIA
calculation becomes repulsive for a kinetic energy of $T_{\rm lab}
\simeq 150-200$~MeV ($|{\bf p'}| \simeq 550-650$~MeV). This implies that
at a momentum transfer of $|{\bf q}|=1$~GeV the distorting potential is
repulsive over the entire range of missing momenta shown in the figure,
thus, producing the observed shift. In Table~\ref{tableone} we display
the sensitivity of the longitudinal asymmetry to distortion effects.
The RDWIA asymmetry is consistently larger than the plane-wave results
due to the added effect of the repulsive distorting potential.

\section{Conclusions}
\label{secconcl}

We have calculated the asymmetry in the missing-momentum distribution
from the $(e,e'p)$ reaction in a relativistic plane-wave impulse
approximation, using a mean-field approximation to the Walecka model.
Longitudinal and transverse nuclear-response functions have been
evaluated in parallel kinematics as a function of the three-momentum
transfer to the nucleus. By dividing out off-shell, single-nucleon form
factors from the response functions, we were able to define a nonrelativistic
plane-wave limit with a symmetric momentum distribution. This provides a
baseline for comparing different theoretical mechanisms of breaking the
symmetry.

In a relativistic formalism the presence of a lower component in the
Dirac bound-state wave function is sufficient to break the symmetry
at the plane-wave level. We have developed analytic expressions for the
plane-wave responses and have illustrated our findings by calculating
the missing-momentum asymmetry using ${}^{16}$O as an example. The
ground state for ${}^{16}$O was generated self-consistently in a
mean-field approximation to the Walecka model. The mean-field
approximation is characterized by the presence of strong (attractive)
scalar and (repulsive) timelike vector potentials. The scalar potential
leads to a reduction of the effective nucleon mass in the medium and is,
ultimately, responsible for a substantial enhancement of the ratio
of upper to lower components in the bound-state wave function.
This dynamical enhancement of the ratio, together with the
increasing importance of the relativistic corrections with momentum
transfer, yield longitudinal asymmetries as large as $A_{L}
\sim 0.8$ for a three-momentum transfer of $|{\bf q}| \sim 10~$GeV.
These nuclear-structure effects preclude the use of the asymmetry in
the missing-momentum distribution as a definitive signature of color
transparency. However, the missing-momentum asymmetry does provide
interesting information about the role of relativity in the nuclear
medium.

We have also investigated the effects of distortions.
We include distortions to assess the impact of the
restoration of gauge invariance on the asymmetry. The RDWIA
calculations yield a larger longitudinal asymmetry than in
the plane-wave case. This results from the repulsive character
of the real distorting potential. We have stressed that a distorting
potential consistent with the mean field used to generate the nuclear
ground state is essential for the conservation of the electromagnetic
current. These mean-field distortions ignore, however, some important
physics -- absorption and energy-dependence -- associated with the
in-medium propagation of the outgoing proton. We believe that a
gauge-invariant treatment of the $(e,e'p)$ reaction that
successfully describes the distortions of the outgoing proton
is an important area for future work.

\acknowledgments

This research was supported by the U.S. Department of Energy
contract \# DE-FG02-87ER40365 (S.G.), as well as by the Florida
State University Supercomputer Computations Research Institute
through the U.S. Department of Energy contracts \# DE-FC05-85ER250000
and \# DE-FG05-92ER40750 (J.P.).

\begin{figure}
 \caption{Charge response in a RPWIA
          for the two valence orbitals in ${}^{16}$O at the
          momentum transfers of $|{\bf q}|=1, 5$~GeV.
          The quantities in brackets give the value
          of the missing-momentum asymmetry.}
 \label{figzero}
\end{figure}
\begin{figure}
 \caption{Longitudinal reduced cross section in a RPWIA
          for the two valence orbitals in ${}^{16}$O at the
          momentum transfers of $|{\bf q}|=1, 5$~GeV.
          The quantities in brackets give the value
          of the missing-momentum asymmetry.}
 \label{figone}
\end{figure}
\begin{figure}
 \caption{Longitudinal missing-momentum asymmetry in a RPWIA
          as a function of the momentum transfer
          $|{\bf q}|$ for the two valence orbitals in ${}^{16}$O.
          The integrals were carried out up to a maximum value
          of the missing momentum of 100~(solid), 200~(dashed),
          and 300 MeV~(dot-dashed), respectively.}
 \label{figtwo}
\end{figure}
\begin{figure}
 \caption{Transverse missing-momentum asymmetry in a RPWIA
          as a function of the momentum transfer
          $|{\bf q}|$ for the two valence orbitals in ${}^{16}$O.
          The integrals were carried out up to a maximum value
          of the missing momentum of 100~(solid), 200~(dashed),
          and 300 MeV~(dot-dashed), respectively.}
 \label{figthree}
\end{figure}
\begin{figure}
 \caption{Longitudinal reduced cross section
          for the two valence orbitals in ${}^{16}$O at a
          momentum transfer of $|{\bf q}|=1$~GeV.
          The dashed(solid) line displays the
          result of a relativistic calculation obtained by
          ignoring(including) distortion effects.
          The dot-dashed line shows the results obtained when
          the lower component of the bound-state wave function in
          dashed calculation is turned off.
          The quantities in brackets give the value
          of the missing-momentum asymmetry.}
 \label{figfour}
\end{figure}

\mediumtext
 \begin{table}
  \caption{Longitudinal numbers of events and
           missing-momentum
           asymmetries as a function of the three-momentum
           transfer with $p_{\rm max}=350~$MeV
           for the $1p^{1/2}$ state in ${}^{16}$O.
           The first(second) row of numbers are the results
           of a relativistic calculation that neglects(incorporates)
           distortion effects.}
   \begin{tabular}{cccc}
    $|{\bf q}|$(GeV) & $N_{L}^{(+)}$ & $N_{L}^{(-)}$ & $A_{L}$ \\
        \tableline
        \tableline
    1.00 & 1.49 & 0.67 &  0.38  \\
         & 1.78 & 0.41 &  0.62  \\
        \tableline
    1.50 & 1.74 & 0.58 &  0.50  \\
         & 1.87 & 0.24 &  0.77  \\
        \tableline
    2.00 & 1.99 & 0.51 &  0.59  \\
         & 2.05 & 0.21 &  0.82  \\
        \tableline
    2.50 & 2.23 & 0.47 &  0.65  \\
         & 2.30 & 0.19 &  0.85  \\
        \tableline
    3.00 & 2.46 & 0.44 &  0.70  \\
         & 2.62 & 0.18 &  0.87  \\
   \end{tabular}
  \label{tableone}
 \end{table}

\end{document}